# A Guide to Teaching Data Science

Stephanie C. Hicks[1,2], Rafael A. Irizarry[1,2]

[1]Department of Biostatistics and Computational Biology, Dana-Farber Cancer Institute, Boston, MA

[2]Department of Biostatistics, Harvard School of Public Health, Boston, MA

Emails:

Stephanie C. Hicks, shicks@jimmy.harvard.edu

Rafael A. Irizarry, rafa@jimmy.harvard.edu

# Abstract

Demand for data science education is surging and traditional courses offered by statistics departments are not meeting the needs of those seeking training. This has led to a number of opinion pieces advocating for an update to the Statistics curriculum. The unifying recommendation is computing should play a more prominent role. We strongly agree with this recommendation, but advocate the main priority is to bring applications to the forefront as proposed by Nolan and Speed (1999). We also argue that the individuals tasked with developing data science courses should not only have statistical training, but also have experience analyzing data with the main objective of solving real-world problems. Here, we share a set of general principles and offer a detailed guide derived from our successful experience developing and teaching a graduate-level, introductory data science course centered entirely on case studies. We argue for the importance of *statistical thinking*, as defined by Wild and Pfannkuck (1999) and describe how our approach teaches students three key skills needed to succeed in data science, which we refer to as *creating*, *connecting*, and *computing*. This guide can also be used for statisticians wanting to gain more practical knowledge about data science before embarking on teaching an introductory course.

# 1. INTRODUCTION

## 1.1 What do we mean by *Data Science*?

The term *data science* is used differently in different contexts since the needs of data-driven enterprises are varied and include acquisition, management, processing, and interpretation of data as well as communication of insights and development of software to perform these tasks. Michael Hochster[1] defined two broad categorizations of data scientists: "*Type A [for 'Analysis'] Data Scientist*" and "*Type B [for 'Building'] Data Scientist*", where Type A is "*similar to a statistician… but knows all the practical details of working with data that aren't taught in the statistics curriculum*" and Type B are "*very strong coders ... may be trained as software engineers and mainly interested in using data 'in production'*". Here we focus on the term *data science* as it refers generally to Type A data scientists who process and interpret data as it pertains to answering real-world questions. We do not make any recommendations as it pertains to training Type B data scientist as we view this as a task better suited for engineering or computer science departments.

## 1.2 Why are statistics[2] departments of a natural home for Data Science in Academia?

Current successful Data Science education initiatives in academia have resulted from combined efforts from different departments. Here we argue that statistics departments should be part of these collaborations. The statistics discipline was born directly from the endeavour most commonly associated with data science: data processing and interpretation as it pertains to answering real world questions. Most of the principles, frameworks and methodologies that encompass this discipline were originally developed as solutions to practical problems.

---

[1] https://www.quora.com/What-is-data-science
[2] We include biostatistics departments

Furthermore, one would be hard pressed to find a successful data analysis by a modern data scientist that is not grounded, in some form or another, in some statistical principle or method. Concepts such as inference, modelling, and data visualization, are an integral part of the toolbox of the modern data scientist. Wild and Pfannkuck (1999) describe *applied statistics* as:

> "*part of the information gathering and learning process which, in an ideal world, is undertaken to inform decisions and actions. With industry, medicine and many other sectors of society increasingly relying on data for decision making, statistics should be an integral part of the emerging information era*".

A department that embraces *applied statistics* as defined above is a natural home for data science in academia. For a larger summary of the current discussions in the statistical literature describing how past contributions of the field have influenced today's data science, we refer the reader to Supplementary Section 1.

**1.3 What is missing in the current Statistics curriculum?** *Creating, Computing, Connecting*

Despite important subject matter insights and the discipline's applied roots, research in current academic statistics departments has mostly focused on developing general data analysis techniques and studying their theoretical properties. This theoretical focus spills over to the curriculum since a major goal is to produce the next generation of academic statisticians. Wild and Pfannkuck (1999) complained that:

> "*Large parts of the investigative process, such as problem analysis and measurement, have been largely abandoned by statisticians and statistics educators to the realm of the particular, perhaps to be developed separately within other disciplines*".

They add that "*[t]he arid, context-free landscape on which so many examples used in statistics teaching are built ensures that large numbers of students never even see, let alone engage in, statistical thinking*".

While academic statisticians have made important theoretical and methodological contributions that are indispensable to the modern data scientist, the theoretical focus does not fully serve the needs of those searching for training in data science. In response to this, numerous authors have offered compelling arguments to expand and update the statistics curriculum emphasizing the importance of data science proficiency. For a discussion on expanding and updating the statistics curriculum, we refer the reader to Supplementary Section 2.

Here, we share our approach to developing and teaching a graduate-level, introductory data science course, which is motivated by an aspiration to emphasize the teaching of three key skills relevant to data science, which we denote as *creating*, *connecting*, and *computing*. These are skills that applied statisticians learn informally during their thesis work, in research collaborations with subject matter experts, or on the job in industry, but not necessarily in traditional courses. We explain each one below and later provide specific advice on approaches that can help students attain these skills.

**Computing:** The topic most discussed in opinion pieces about updating the curriculum is the lack of computing. For example, Nolan and Temple Lang (2012), Speed (2014), and Baumer (2015) all argue that there should be more computing in the statistics curriculum. These opinions ring true for us as researchers in the field of genomics in which computing is an essential skill. This is particularly the case for those of us that disseminate our statistical methodology by developing and sharing open source software (Huber et al. 2015).

**Connecting:** In our view, computing is not the main disconnect between the current statistics curriculum and data science. Instead, we contend the main problem is that traditional statistics courses have focused on describing techniques and their mathematical properties rather than solving real-world problems or answering questions with data. Data are often included only as examples to illustrate how to implement the statistical technique. While knowledge and understanding of the techniques are indispensable to using them effectively and to avoid reinventing the wheel, the more relevant skill pertaining to data science is to **connect** the subject matter question with the appropriate dataset and analysis tools, which is not currently prioritized in statistics curricula. This relates to what happens during what Wild and Pfannkuck (1999) refer to as the *interrogative cycle* of *statistical thinking.* In contrast, the typical approach in the classroom is to mathematically demonstrate that a specific method is an optimal solution to something, and then illustrate the method with an unrealistically clean dataset that fits the assumptions of the method in an equally unrealistic way. When students use this approach to solve problems in the real-world, they are unable to discern if the dataset is even appropriate for answering the question and unable to identify the most appropriate methodological approach when it is not spoon fed. Furthermore, knowing how to leverage application-specific domain knowledge and interactive data visualization to guide the choice of statistical techniques is a crucial data science skill that is currently relegated to learning on the job. For a much more detailed exposition on this idea, consult Wild and Pfannkuck (1999).

**Creating:** The current statistical curriculum tacitly teaches students to be passive: wait for subject-matter experts to come to you with questions. However, in practice data scientists are often expected to be active; they are expected to **create** and to formulate questions. Note that many of the founders of the discipline of Statistics did just this. For example, as early as the 1830s, Quetelet created large demographic datasets to describe, for example, the association between crime and other social factors (Beirne 1987). Francis Galton collected father and son height data to answer questions about heritability (Galton 1886 and 1889). Ronald Fisher designed agricultural experiments and collected yield data to determine if fertilizers helped crops (Fisher 1921). Examples abound, but in today's academia this is much more common among data analysts in other departments such as Economics where, for example, data analysts are designing randomized controlled trials, which are increasingly being used to guide policy interventions[3].

## 1.4 Bridging the gap in the classroom to teach Data Science

While many have called for statisticians to rebuild the undergraduate curriculum (Cobb 2015) or that statistics departments build data science courses (Baumer 2015; Hardin et al. 2015), here, we add that the individuals tasked with bridging the gap must be experienced themselves in *creating*, *connecting* and *computing*. Adding data and computing to a standard course is not sufficient. It should not be assumed that once data appears, theoretical knowledge automatically translates into the necessary skill to build an effective course or curriculum. For the very same limitations with the curriculum described above, a PhD in Statistics is not sufficient. We therefore *encourage* applied statisticians experienced in creating, connecting and computing to become *involved* in the development of courses and updating curricula. Similarly, we encourage statistics departments to reach out to practicing data analysts, perhaps in other departments or from other disciplines, to collaborate in developing these courses.

---

[3] Poverty Action Lab (https://www.povertyactionlab.org/), Government Performance Lab (http://govlab.hks.harvard.edu/), and Education Innovation Laboratory (http://edlabs.harvard.edu/)

To aid in bridging the gap, as applied statisticians that have taught data sciences courses, we synthesize our recommendation on how to prepare and teach a graduate-level, introductory course in data science. We share a set of guiding principles and offer a detailed guide on how to teach an introductory course to data science. This guide can also be used for statisticians wanting to gain more practical knowledge and experience in computing, connecting and creating before embarking on teaching a data science course. In the Discussion Section, we include thoughts and recommendations for more advanced or specific data science courses.

## 2. PRINCIPLES OF TEACHING DATA SCIENCE

The key elements of our course can be summarize into five guiding principles (Table 1).

| **Principles of Teaching Data Science** |
|---|
| ● Organize the course around a set of diverse case studies<br>● Integrate computing into every aspect of the course<br>● Teach abstraction, but minimize reliance on mathematical notation<br>● Structure course activities to realistically mimic a data scientist's experience<br>● Demonstrate the importance of critical thinking/skepticism through examples |

**Table 1.** A set of general principles of teaching data science meant to be used as a guide for individuals developing a data science course.

### 2.1. Organize the course around a set of diverse case studies

The first and most fundamental change we made to the standard statistical course pertaining to data science was to bring the subject matter question to the forefront and treat the statistical techniques and computing as tools that help answer the question. Through answering these questions, students learn to connect subject matter to statistical frameworks and also learn new statistical techniques as part of the data analysis process. Only after the techniques are motivated and introduced as solutions to a problem, are the mathematical details and justifications described. We carefully and deliberately select a set of diverse case studies to give us an opportunity to motivate and teach the wide range of techniques needed to be a successful data scientist. We discuss specific case studies in Section 3.2.

This approach is certainly not new and was explicitly proposed in 1999 by Nolan and Speed in the book *Stat Labs*, which promotes teaching mathematical statistics through in-depth case studies (Nolan and Speed 1999 and 2000). The abstract of the book eloquently describes the main guiding principle we followed when developing the class:

> *"Traditional statistics texts have many small numerical examples in each chapter to illustrate a topic in statistical theory. Here, we instead make a case study the centerpiece of each chapter. The case studies, which we call labs, raise interesting scientific questions, and figuring out how to answer a question is the starting point for developing statistical theory. The labs are substantial exercises; they have nontrivial solutions that leave room for different analyses of the data. In addition to providing the framework and motivation for studying topics in mathematical statistics, the labs help students develop statistical thinking. We feel that this approach integrates theoretical and applied statistics in a way not commonly encountered in an undergraduate text."*

In a related recent blog post[4], Jeff Leek describes the need to put the *problem first.* We note that in our experience selecting motivating case studies and developing an illustrative data analysis for each one required the largest time investment.

**2.2. Integrate computing into every aspect of the course**

Donald Knuth originally introduced the idea of literate programming, which weaves instructions, documentation and detailed comments in between machine-executable code producing a document that describes the program that is best for human understanding (Knuth 1984). Recently, educational and scientific communities have begun to adopt this concept in the classroom by increasing the use of computing and requiring student assignments be completed using literate programming (Baumer et al. 2014). Here, we also advocate that the lectures in a data science course also be created with literate programming. This demonstrates to students what is expected in the homework assignments, stresses the importance of computational skills (Nolan and Temple Lang 2012), and maximizes the "hands-on" experience of data analysis in classroom. This allows students to observe the code used by an expert, the instructor. Furthermore, it also permits the instructor to write code live and to explore data during lecture, which gives students an opportunity to see and to hear an expert's rationale for the choice of statistical technique and computational tool.

The use of literate programming also promotes reproducible research (Claerbout 1994; Buckheit and Donoho 1995). In the classroom, this is useful not only for the student assignments and projects, which provides a practical learning experience, but also the lectures, because it can alleviate a major frustration for students who might spend a significant time outside of class trying to repeat the steps of an analysis presented in lecture that did not integrate the data, code and results in one document.

---

[4] http://simplystatistics.org/2013/05/29/what-statistics-should-do-about-big-data-problem-forward-not-solution-backward/

Finally, literate programming also facilitates the implementation of active learning approaches. Given that students need to absorb subject matter knowledge, statistical concepts, and computing skills simultaneously (and falling behind on any one can result in a student missing out key concepts), we find that regularly challenging students with assessments permits the lecturer to gauge adoption and adapt accordingly. Section 3.3. provides more details on how we implemented this principle.

We note that computing is a rather complex topic within itself. Programming requires ways of thinking that are difficult to develop. In our course, we assume that students have some programming skills. These skills can be attained in a basic programming course including several free online courses.

**2.3. Teach abstraction, but minimize reliance on mathematical notation**

Abstraction in the context of data analysis is one of the statistical discipline's most important contributions. We agree with Wild and Pfannkuck that "*[t]he cornerstone of teaching in any area is the development of a theoretical structure with which to make sense of experience, to learn from it and transfer insights to others.*" Although we do not implement the standard approach to teaching abstraction, we do consider it a fundamental aspect of data science and therefore recommend teaching it. One way in which we differ from the standard approach is that we avoid mathematical notation when possible. Mathematical notation provides the most efficient way of describing statistical ideas. However, we find that it often obfuscates the teaching of fundamental concepts. So when possible, we use computational approaches, such as data visualization or Monte Carlo simulations (Cobb 2007; Hesterberg 2015; Horton 2015) to convey concepts. Minimizing mathematical notation also forces students to grasp the ideas in a more intuitive way. George Cobb refers to this as "seeking depth" or taking away what is technical (formalisms and formulas) to reveal what is fundamental (Cobb 2015). Furthermore, mathematical formalisms are easier to learn after concepts have been grasped.

**2.4. Structure course activities to realistically mimic a data scientist's experience**

Once a question has been formulated, the data scientist's experience can be described as an iterative and interactive process which involves learning about the subject matter, data wrangling, exploratory data analysis, and implementing or developing statistical methodology. Wild and Pfannkuck (1999) describe "s*tructured frameworks*" for this process using the *Problem, Plan, Data, Analysis, Conclusion* cycle which is "concerned with abstracting and solving a statistical problem grounded in a larger 'real' problem.... *Knowledge gained and needs identified within these cycles may initiate further investigative cycles*". In this process it is rarely clear what method is appropriate, nor is there an obvious, uniquely correct answer or approach. Mimicking this experience will greatly help students learn to **connect**. Homeworks should be designed to be open-ended and based on real-world case studies. Students should be given an opportunity to commence work before the statistical techniques or software tools, considered to be appropriate by the instructor, are introduced. Lectures should demonstrate how to efficiently use **computing** in the data analysis process and homeworks give students time to practice and become proficient at that skill. In a final project, students can be asked to define their own question, which will help them learn to **create**.

## 2.5. Demonstrate the importance of critical thinking/skepticism through examples

As Tukey[5] once said "*[t]he combination of some data and an aching desire for an answer does not ensure that a reasonable answer can be extracted from a given body of data.*" Culturally, academic statisticians are cautious and skeptical. Today's data science culture is much more optimistic about what can be achieved through data analysis. Although we agree that overly pessimistic attitudes should be avoided as they can lead to missed opportunities (Speed 2014), we are mutually concerned that there is currently too much hubris. To that end, we would argue that it is indispensable that data science courses underscore the importance of critical thinking and skepticism. Being skeptical and cautious has actually led to many important contributions. An important example is how randomized controlled experiments changed how medical procedures are evaluated (Freedman, Pisani and Purves 1998). A more recent one is the concept of the false discovery rate (Benjamini and Hochberg 1995), which has helped reduce the reporting of false positives in, for example, high-throughput biology (Storey and Tibshirani 2003). Ignoring these contributions has led to analysts to being fooled by confounders, multiple testing, bias, and overfitting[6]. Wild and Pfannkuck (1999) point out that "[a]s you gain experience and see ways in which certain types of information can be unsoundly based and turn out to be false, you become more sceptical". We agree and the only way we know how to teach this is through illustration with real examples. For example, providing case-studies of Simpson's Paradox can illustrate how confounding can cause spurious associations (Horton 2015).

---

[5] http://www.azquotes.com/quote/603406
[6] http://simplystatistics.org/2014/05/07/why-big-data-is-in-trouble-they-forgot-about-applied-statistics/

# 3. CASE STUDY: INTRODUCTION TO DATA SCIENCE

We developed and taught a graduate course titled Introduction to Data Science (BIO 260) at the Harvard T.H. Chan School of Public Health (HSPH) in Spring 2016 (http://datasciencelabs.github.io/). Here, we present this course as a demonstration with specific examples of how we implemented our principles of teaching an introductory course in data science.

**3.1 Course logistics**

We used the R programming language, but we have used Python in another similar course (http://cs109.github.io/2014/). We do not consider the choice between R and Python an important one, but we do find beginers have an easier time learning R, in particular when the R packages in the *tidyverse* (Wickham 2016), such as *dplyr* (Wickham and Francois 2016) and *ggplot2* (Wickham 2009), are used.

Due to the effort required to grade homeworks, which included open-ended questions, we had at least one TA per 15 students, but we would recommend an even higher TA to student ratio. This high TA to student ratio was key to the success of the course as grading open-ended homeworks was highly time consuming.

The course assumes that students have some programming experience and knowledge of statistics at an introductory level. There were no formal textbooks or reading requirements. Students had diverse educational backgrounds and interests (Figure 1). This diversity presented a challenge as most of the material was review for some and difficult for others. The high TA to student ratio helped those that needed review and having the lectures posted before class helped the students determine if they needed to attend the lecture that data. We also made videos of the lectures and code used in class available for students to review.

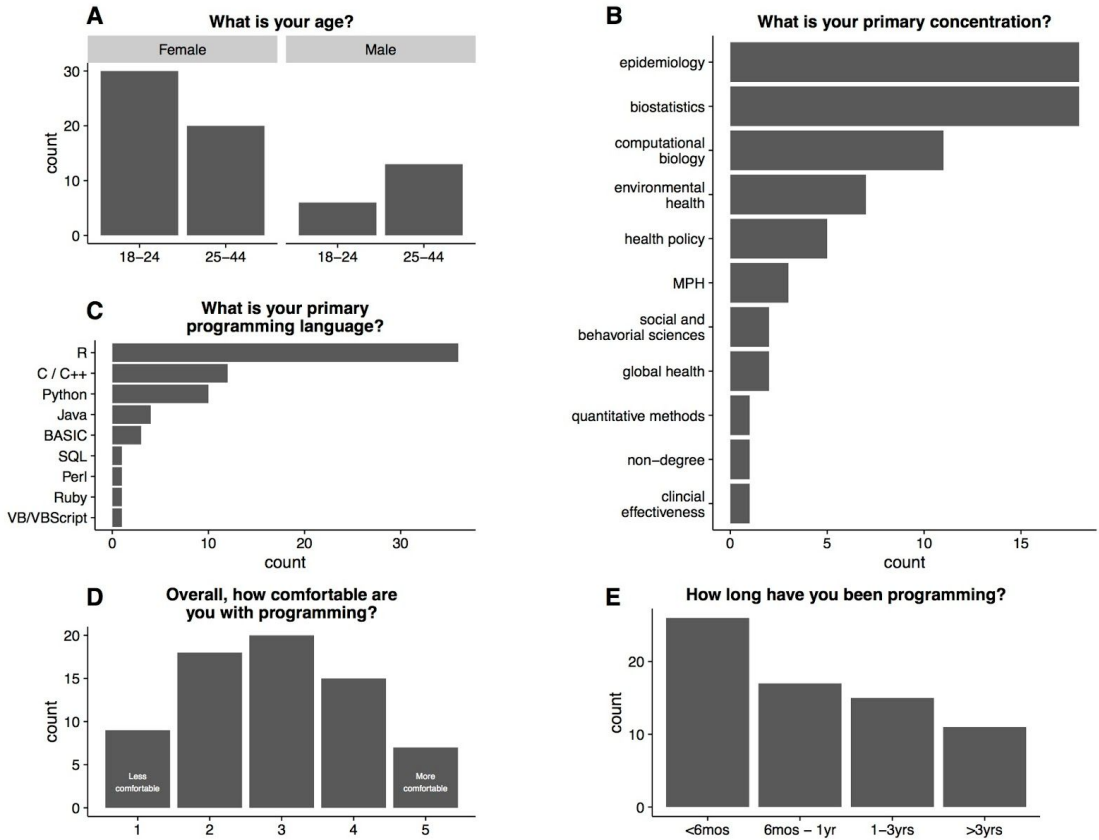

**Figure 1.** Summary from self-reported survey of students in the Introduction to Data Science (BIO 260) course taught at Harvard T.H. Chan School of Public Health in Spring 2016.

**3.2 Case Studies**

As described in Section 2.1, the most important feature of this course is that we brought the subject matter question to the forefront. We structured the course and demonstrated the entire data analysis process using a set of case studies (Table 2). We selected the case studies from diverse applications and of interests to our students. These case studies were also chosen to assure that a variety of techniques were motivated to be useful tools to answer the questions. These included summary statistics, exploratory data analysis, data wrangling, inference, modeling, regression, Bayesian statistics, databases, and machine learning. The "*Two Cultures*" as described by Breiman (2001) received equal time with the course with equal time between exploratory data analysis/visualization, inference/modeling, and machine learning. In Supplemental Section 3, we describe seven case studies that occupied most of the semester. Other topics we covered can be see in detail on the course website (http://datasciencelabs.github.io/pages/lectures.html).

To prepare a given case study, we used the following steps: (1) Decide what concepts or skills (column 3 of Table 2) we want to cover. (2) Look for example questions that are of general interest (column 1 of Table 1). (3) Identify a public dataset (column 2 of Table 1) that we can use to answer the question. (4) Answer the question ourselves and summarize it into an R markdown report. (5) Determine if the solution is pedagogically acceptable (e.g. not too hard or too easy). (6) Separate material to be covered in class from material to be left for the homework.

| Motivating question | Data set | Concepts/Skills learned | | Number of lectures |
|---|---|---|---|---|
| How tall are we? | Self-reported heights of students in course | Introduction to R, R Markdown, RStudio, Exploratory data analysis (EDA), summary statistics | | 3 |
| Are health outcomes and income inequality increasing or decreasing in the world? | Global health and economic data from Gapminder | Exploratory data analysis; Static and interactive data visualization using ggplot2 | | 4 |
| What are the basics of data wrangling? | Sleep times and weights for a set of mammals | Data wrangling using dplyr, tidyr | | 1 |
| How do I get the course materials and submit homework? | N/A | Version control with git/GitHub | | 1 |
| Who is going to win the 2016 Republican Primary? | U.S. Opinion polls from the 2008 and 2012 Presidential Election, 2016 Presidential Primaries | Probability, statistical inference Monte Carlo simulation; smoothing; Introduction to models, Bayesian statistics | | 5 |
| Midterm #1 | | | | 1 |
| How to pick players for a baseball team with a limited budget? | Sabermetrics baseball data | Regression; Data wrangling using broom | | 3 |
| How to predict handwritten digits? | MNIST database of handwritten digits | Dimension reduction; Machine learning; Cross validation; Performance evaluation | | 4 |
| How can I scrape data from a website? | Amazon book reviews for *ggplot2: Elegant Graphics for Data Analysis (Use R!)* | Scraping data from web | Using rvest/CSS Selectors; regular expressions | 1 |
| How can I get data using an API? | Tweets with #rstats hashtag from Twitter | | APIs; Sentiment analysis | 1 |
| Will I like this movie? | MovieLens user ratings of movies | Regularization; Matrix Factorization | | 1 |
| Can we identify power plants within 10km of an air pollution monitor? | Power plants locations and measured air pollution | Structured Query Language (SQL); Relational databases | | 1 |
| Is there gender bias in grant funding? | PNAS | Confounding; Simpson's Paradox | | 1 |
| Midterm #2 | | | | 1 |
| Final Project Presentations | | | | 1 |

**Table 2.** A table describing the case studies in our Introduction to Data Science course. Each case study is defined by a motivating problem or question, the data set used and the concepts and skills learned. We also provide the number of lectures (29 total) corresponding to the time spent on each topic.

We note that not all examples selected in Step 2 satisfied all six steps and were thus discarded. For this reason, we had to consider several different examples in Step 2. We estimate that this process took an average of 40 hours per case study.

**3.3 Integrate computing in all aspects of the course**

*Literate Programming: Using R Markdown and R pres*

Each lecture and homework assignment was created using literate programming. We prepared lectures using R Markdown (Rmd) and R Presentations (Rpres) and rendered the presentations using RStudio, which provides functionality to easily convert from these formats to PDF or HTML (Xie 2015). More importantly, using RStudio also permitted us to run live data analysis during lecture. These documents were available on GitHub (https://github.com/datasciencelabs/2016) to allow students to follow along and run code on their own laptops during class. For each lecture, there were three to four TAs available in the classroom who were walking around to answer questions in person. In addition, we included a link to a Google Document at the top of the R Markdown in each lecture to allow students a venue to ask questions if they did not want to interrupt the lecture. Note that in the course in which we used Python, we used Jupyter Notebooks which provide similar functionality to Rmd and Rpres. Karl Broman has provided several useful tutorials[7] in these formats and others.

*Using active learning techniques to teach data science*

We divided lectures into 10 to 30 minutes modules and included 3-5 assessment problems in between. These questions consisted of multiple-choice or open-ended questions with most of them requiring a short data analysis. The solutions, in the form of code required to solve these assessments, were presented and discussed in class and added to the lectures only after the lecture was complete. We asked students to enter their answers in Google Forms that we created before lecture (Figure 2). Seeing these responses permitted us adapt the pace of the lectures.

---

[7] http://kbroman.org/pages/tutorials.html

### 3.4 Structure course activities to realistically mimic a data scientist's experience

**Lectures and homeworks**

Each lecture and homework was structured to mimic a data scientist's experience by analyzing a data set as an in depth case study. Out of the six homework assignments, five were directly connected to the case studies discussed in the lecture. Students were provided the assignments prior to the corresponding lecture to self-familiarize themselves with the details of the real-world problem and promote initial thoughts and data explorations. During class, lectures began with a formal introduction to the real-world problem immediately followed by a discussion on how to **create** and actively formulate a question that can be answered with data as it relates to the problem at hand. Once a question had been formulated, the data, course material, code, and analysis tools were introduced and woven into the lecture in the order that they were needed to demonstrate the iterative and interactive data analysis process. This approach demonstrated to the students how to **connect** the subject matter question with the appropriate data and analysis tool. Furthermore, this approach emphasized the importance of **computing** because students could run the code provided in the lecture on their own laptops live in class.

Because the homework assignments were interconnected to the lectures (extensions of the same in depth case studies discussed in lecture), the students could immediately dive into the data analysis, answering both determinate and open-ended questions. The former provided a concrete and consistent way to grade the assignments and the latter promoted creative freedom in the choice of what techniques to apply and demonstrated the iterative data process, similar to a data scientist's experience.

**Midterm exams**

During the majority of assessments, students worked with TAs and other students to complete assignments. To assure that each individual student was grasping basic concepts and gaining basic skills, we incorporated two midterm exams (one halfway and one at the end) in the course.

**The Final Project**

The students also completed a month long final project on a topic of their choice either on their own or in a group. This portion of the course most closely mimicked the data scientist's experience. The deliverables for the project included a project proposal, a written report of the data analysis in an R Markdown, a website, and a two-minute video communicating what the group learned. The project proposal described the motivation for the project, the project objectives, a description of the data, how to obtain the data, an overview of the computational methods proposed to analyze the data and a timeline for completing the project. TAs were paired together with 3-4 groups to meet to discuss the proposed projects and provide guidance. The students used the concepts learned in and outside of the course to complete the projects. Once the projects were complete, the submitted deliverables were reviewed and the best projects were highlighted at the end of the course.

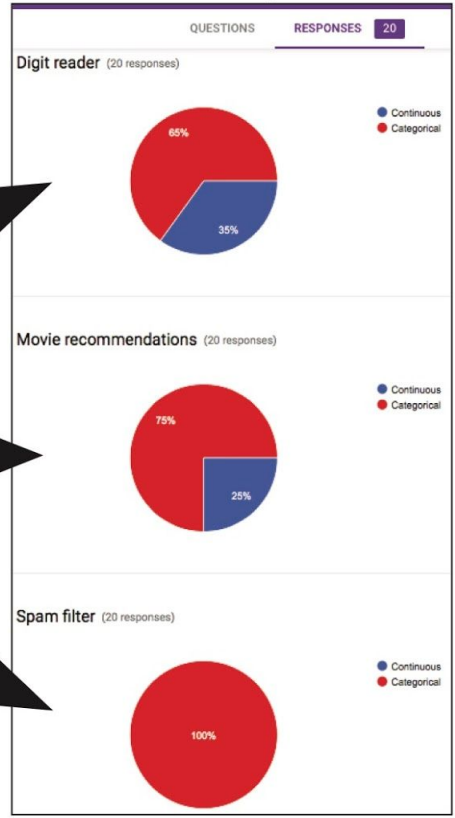

**Figure 2.** Using Google Forms as an active learning tool. (A) Three to five assessments were included in the R Markdown for each lecture, which consisted of either multiple-choice or open-ended questions. (B) Students were given a few minutes during the lecture to answer the questions. (C) Student responses were recorded and instructors could see the responses instantly. The live responses helped adapt the pace of the lectures.

**3.5 Organizing Course Content with GitHub**

We used git and GitHub repositories to store course content, build the course website, and organize homework submissions. GitHub provides a service, GitHub Pages (https://pages.github.com/), that facilitates the construction of the website (http://datasciencelabs.github.io/). The course content (https://github.com/datasciencelabs/2016) and data (https://github.com/datasciencelabs/data) were available in GitHub repositories (Figure 3). Note that on github we organized the material by concept and skill rather than by case study. We did this because it permitted students to look-up concepts during their final project.

Although we could have organized the course with a standard teaching tool such as Blackboard or Canvas, we chose to expose students to the notion of version control and achieved this by using one of the most popular systems, git, along with the web-based git repository hosting service, GitHub. GitHub is currently the most widely used resource for code developers including data scientists.

We spent one lecture introducing the concept of version control, git and GitHub. To demonstrate the mechanics, we created a test repository (https://github.com/datasciencelabs/test_repo) and asked all the students to use git to obtain a copy of this repository during the lecture. We also introduced the concept of making changes to local repositories and pushing the changes to remote repositories. After this lecture, students were able to stay in sync with the course repository to access the course material at the beginning of each lecture.

In addition, we used git and GitHub to create private repositories for students to submit their homework assignments (Figure 3). GitHub Classroom ([https://classroom.github.com/](https://classroom.github.com/)) offers free private repositories to instructors, which can be used to create assignments and distribute starter code to students. Each student was given access to make changes to his or her own private repository to submit their homework assignment. The last commit was used as the final submission. The TAs were able to quickly and efficiently access and grade the homework submissions. Note that GitHub regularly offers new services so we recommend keeping up to date with the latest.

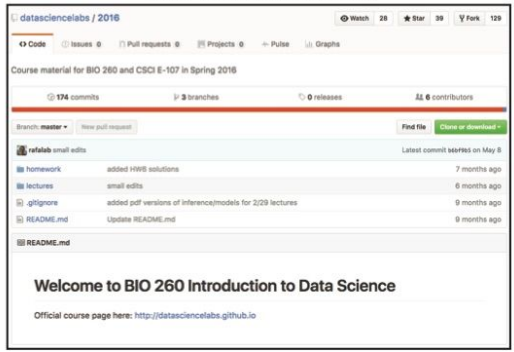
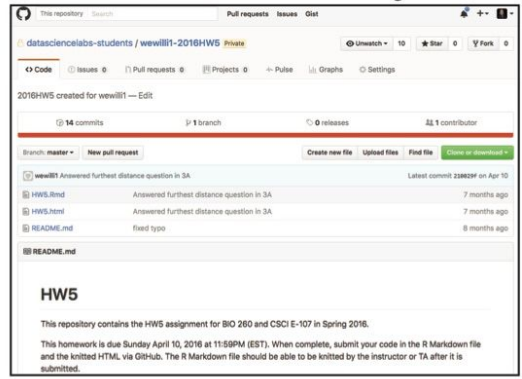
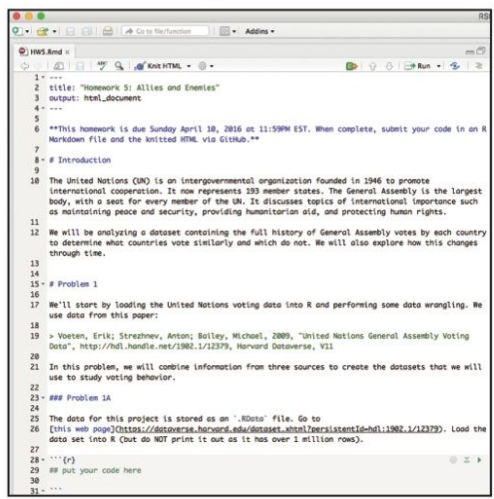
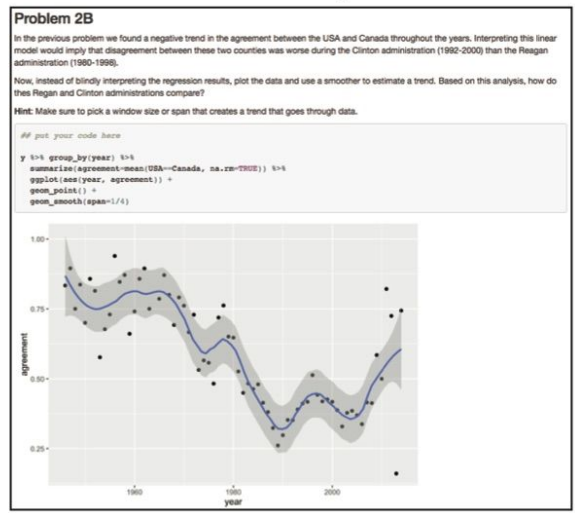

**Figure 3.** Using git and GitHub to organize course materials and submit homework. (A) The course website was built with GitHub pages and a public course repository was created to incorporate the course material (lectures, homework, solutions and data). (B) Private GitHub repositories were created for each student and each homework assignment that contained starter code with the homework assignment. (C) Homework assignments were created in R Markdown and specific code chunks were created for the students to add their code as solutions. (D) Once the student was satisfied with their solutions, the homework submission was committed to the private GitHub repository as an R Markdown and HTML. The TAs were able to quickly and efficiently access and grade the homework submissions in the individual repositories.

# 4. DISCUSSION

We have presented a set of principles and a detailed guide for teaching an introductory data science course. The course was centered around a set of case studies and homeworks that permitted students an experience that closely mimicked that of a data scientist's. The case studies were carefully chosen to motivate students to learn important statistical concepts and computing skills. Choosing these case studies, finding relevant data sources, and preparing didactic data analysis solutions required an enormous initial investment of time and effort. We recommend that academic institutions invest accordingly on faculty preparing these courses by, for example, increasing the percent effort associated with teaching theses courses for the first time (Waller 2017). At institutions that can't accommodate this, it may be necessary to form interdepartmental teams to assure that enough faculty teaching time is allocated to the course. Going forward, we will need new case studies and we propose that, as a community of educators, we join forces to create a collection of case studies.

We have taught a version of our introductory course twice: once as an upper-level undergraduate class at Harvard College and once as a graduate course in the School of Public Health. In both cases we considered the course to be a success. In both courses we had a diverse group of students in terms of their educational backgrounds, programming experience and statistical knowledge. We found that students were motivated and willing to do extra work to catch up in their areas of weakness. We attribute this partially to the fact that students were highly motivated to learn the material. Neither of our courses were requirements, thus all students chose to take the class.  We do recognize that our population of students are likely above average in their adeptness to self learning. In fact, through the extension school, the course was offered to a more general population of students, and this population did find the course more challenging. In particular, they required more assistance with statistical concepts and learning Python/R. In future courses, we will set different requirements for different population of students. For example, graduate students will be expected to learn on their own, while extensions school students will be required to know the programming language in advance and have taken an introductory statistics and probability course.

We do note that an even more introductory version of our class can serve as a first course in statistics. Exposing students to practical questions and data in the real-world can serve as great motivation to learning the concepts and mathematics presented in an introductory statistics course. We also note that our course is only an introduction and that several more advanced courses are necessary to meet the demand of those seeking data science training. Under the current makeup of undergraduate curricula, curricula for advanced students will need to flexible. For example, students with strong programming skills, but no statistical training will benefit from courses teaching advanced probability, inference and methods, while statistics with no computing experience may benefit from courses on optimization, algorithms and data structures.

We may also need to develop new courses, or adapt existing ones, that demonstrate real-world examples and applications of (1) Machine Learning, (2) software engineering courses that teach students the principles needed to plan, organize, code, and test data analysis software, (3) advanced computing courses necessary to deal effectively with datasets that do not fit in memory and algorithms that require multiple compute nodes to be implemented in practice, and (4) subject-matter centered courses that expose students to the nuances of data science in the context of specific research areas such the social sciences, astronomy, ecology, education, and finance. Clearly, under current academic structures, data science programs will be best served by collaboration spanning more than one department. All students will benefit from courses putting a greater emphasis on data and data analysis. We are confident that our principles and guide can serve as a resource not just for specific new courses, but also to generally improve masters and PhD level curricula as well.

## Acknowledgments

We thank Joe Blitzstein and Hanspeter Pfister, the creators of CS109, from which we borrowed several logistical ideas, David Robinson, our *tidyverse* and *ggplot2* guru, for advice and for his guest lectures, Alyssa Frazee who helped develop the movie ratings lecture, Joe Paulson for suggesting Google polls, Héctor Corrada-Bravo for advice on teaching Machine Learning, GitHub Education for providing free private repositories, Garrett Grolemund, Sherri Rose and Christine Choirat for presenting guest lectures, all of the TAs from our Introduction to Data Science course BIO260 (Luis Campos, Stephanie Chan, Brian Feeny, Ollie McDonald, Hilary Parker, Kela Roberts, Claudio Rosenberg, Ayshwarya Subramanian) and GroupLens for giving us permission to adapt and re-distribute part of the 'ml-latest' MovieLens data set on our course website. We thank Jeff Leek for comments and suggestions that improved the manuscript and Scott Zeger for a helpful discussion. Finally, we thank NIH R25GM114818 grant for partial support for creating the teaching materials.

# References

Baumer, B., Cetinkaya-Rundel, M., Bray, A., Loi, L., and Horton, N. J. (2014), "R Markdown: Integrating A Reproducible Analysis Tool into Introductory Statistics," *Technology Innovations in Statistics Education*, 8.

Baumer, B. (2015), "A Data Science Course for Undergraduates: Thinking With Data," *The American Statistician*, 69, 334–42.

Beirne, P.. 1987. "Adolphe Quetelet and the origins of positivist criminology". *American Journal of Sociology*, 92, 1140-1169.

Benjamini, Y., Hochberg, Y. (1995), "Controlling the False Discovery Rate: a Practical and Powerful Approach to Multiple Testing," *Journal of the Royal Statistical Society. Series B (Methodological)*, 57, 289-300.

Breiman, L. (2001), "Statistical Modeling: The Two Cultures," *Statistical Science*, 16, 199-231.

Buckheit, J. B., Donoho, D. L. (1995), "WaveLab and Reproducible Research," In *Wavelets and Statistics*, edited by Anestis Antoniadis and Georges Oppenheim, 55–81. Lecture Notes in Statistics 103. Springer New York.

Claerbout, J. (1994), "Hypertext Documents about Reproducible Research," Tech. rep., Stanford University, Available at *http://sepwww.stanford.edu/sep/jon/nrc.html*.

Cobb, G. W. (2015), "Mere Renovation Is Too Little Too Late: We Need to Rethink Our Undergraduate Curriculum from the Ground Up," *The American Statistician*, 69, 266–82.

Cobb, G. W. (2007), "The Introductory Statistics Course: A Ptolemaic Curriculum?", *Technology Innovations in Statistics Education* 1, Available at *https://escholarship.org/uc/item/6hb3k0nz.pdf.*

Freedman, D., Pisani, R., Purves, R. (1998), *Statistics*, W. W. Norton & Company Inc. New York, New York. Chapter 1.

Fisher, R. A. (1921), "Studies in crop variation. I. An examination of the yield of dressed grain from Broadbalk," *Journal of Agricultural Science*, 11, 107-135.

Galton, F. (1886), "Regression Towards Mediocrity in Hereditary Stature," *Journal of the Anthropological Institute of Great Britain and Ireland*, 15, 246-263.

Galton, F. (1889), *Natural Inheritance*, London, Macmillan.

Hardin, J., Hoerl, R., Horton, N. J., et al. (2015), "Data Science in Statistics Curricula: Preparing Students to 'Think with Data'," *The American Statistician*, 69, 343–53.

Hesterberg, T. C. (2015), "What Teachers Should Know About the Bootstrap: Resampling in the Undergraduate Statistics Curriculum," *The American Statistician*, 69, 371–86.

Horton, N. J. (2015), "Challenges and Opportunities for Statistics and Statistical Education: Looking Back, Looking Forward," *The American Statistician*, 69, 138–45.

Horton, N. J., Baumer, B., Wickham, H. (2015), "Setting the Stage for Data Science: Integration of Data Management Skills in Introductory and Second Courses in Statistics," *Chance*, 28: 40–50. Available at *http://chance.amstat.org/2015/04/setting-the-stage/.*

Huber, W., Carey, V. J., Gentleman, R., et al. (2015), "Orchestrating High-throughput Genomic Analysis with Bioconductor," *Nature Methods*, 12, 115-21.

Knuth, D. E. (1984), "Literate Programming," *Computer Journal*, 27, 97–111.

Nolan, D., Temple Lang, D. (2012), "Computing in the Statistics Curricula," *The American Statistician*, 64, 97-107.

Nolan, D., Speed, T. P. (1999), "Teaching Statistics Theory through Applications," *The American Statistician*, 53, 370–75.

Nolan, D., Speed, T. P. (2000), *Stat Labs: Mathematical Statistics Through Applications*, Springer Texts in Statistics. Springer New York.

Speed, T. (2014) "Trilobites and Us," *Amstat News*, American Statistical Association, January 1, Available at *http://magazine.amstat.org/blog/2014/01/01/trilobites-and-us/comment-page-1/*.

Storey, J. D., Tibshirani, R. (2003), "Statistical significance for genomewide studies", *Proceedings of the National Academy of Sciences*, 100, 9440–9445.

Waller, L.A. (2017), "Documenting and Evaluating Data Science Contributions in Academic Promotion in Departments of Statistics and Biostatistics". *bioRxiv*. https://doi.org/10.1101/103093.

Wickham, H. (2009), *Ggplot2: Elegant Graphics for Data Analysis*. New York: Springer.

Wickham, H. (2016), "tidyverse: Easily Install and Load 'Tidyverse' Packages". R package version 1.0.0. Available at *https://CRAN.R-project.org/package=tidyverse*

Wickham, H., Francois, R. (2016), *dplyr: A Grammar of Data Manipulation*. R package version 0.5.0.

Wild, C. J., Pfannkuch, M. (1999), "Statistical Thinking in Empirical Enquiry," *International Statistical Review*, 67, 223-265.

Xie, Y. (2015), *Dynamic Documents with R and Knitr, Second Edition*. Chapman & Hall/CRC The R Series. CRC Press.

# Supplementary Material

**1. How past contributions from Statistics have influenced today's data science**

As summarized by Donoho, "[i]nsightful statisticians have for at least 50 years been laying the groundwork for constructing [Data Science] as an enlargement of traditional academic statistics" (Donoho 2015). For example, consider this insightful quote from John Tukey:

*"For a long time I have thought I was a statistician, interested in inferences from the particular to the general. But as I have watched mathematical statistics evolve, I have had cause to wonder and to doubt. ... All in all I have come to feel that my central interest is in data analysis, which I take to include, among other things: procedures for analyzing data, techniques for interpreting the results of such procedures, ways of planning the gathering of data to make its analysis easier, more precise or more accurate, and all the machinery and results of (mathematical) statistics which apply to analyzing data"* (Tukey 1962).

In fact, in 1998, C.F. Jeff Wu delivered a lecture titled "Statistics = Data Science?" and suggested that that statistics be renamed "data science" and statisticians "data scientists" (Wu 1998). Wild and Pfannkuck (1999) describe *applied statistics* as:

*"part of the information gathering and learning process which, in an ideal world, is undertaken to inform decisions and actions. With industry, medicine and many other sectors of society increasingly relying on data for decision making, statistics should be an integral part of the emerging information era"*.

## 2. Expanding and updating the statistics curriculum

Numerous authors have offered compelling arguments to expand and update the statistics curriculum. For example, in 2014 the American Statistical Association presented a set of curriculum guidelines for undergraduate programs with four key points: increased importance of data science, real applications, more diverse models and approaches and ability to communicate (American Statistical Association Undergraduate Guidelines Workgroup 2014). In 2015, *The American Statistician* published a special issue emphasizing the importance of data science proficiency in the undergraduate statistics curriculum (Horton and Hardin 2015). The issue included a series of articles calling for statisticians to rebuild the undergraduate curriculum (Cobb 2015) and describing data science courses implemented in statistics departments (Baumer 2015; Hardin et al. 2015), consulting courses (Smucker and Bailer 2015; Khachatryan 2015) and workshops (Nolan and Temple Lang 2015) exposing students to modern applied statistical research and practice.

## 3. Description of the case studies

### *How tall are we?*

On the first day of the class we asked students to describe the heights of the members of our class. This starts a conversation about the ways to summarize data including numerical summary statistics and graphical summaries using plots. Students quickly realize we can collect data to answer this question. To organize this process, we let students self-report their height in inches by filling out a survey using Google Forms (Supplemental Figure 1). Eager to commence visualizing the data and computing summaries, students are motivated to learn (1) basic R syntax, (2) how to use RStudio to write, run and view code in reproducible documents (R

Markdown) in an integrated development environment (IDE), and (3) how to import the data into R. After this point, the students are ready for their first lesson in R data types and syntax.

As students attempt to import the data, they have their first experience with the type of challenges faced by data scientists not described in introductory statistics courses, because a substantial number of students did not report their heights in inches or in a consistent manner. For example, some of the heights were not even numerical, '5'10''' or '5 feet 10 inches'. Then, we introduce the concept of data wrangling and describe several available tools to facilitate the conversion from free text to numeric values. We also used this data to introduce exploratory data analysis as a tool for data wrangling to summarize data and identify potential problems (e.g. to detect outliers such as entries in centimeters).

Once the data was successfully loaded and wrangled we introduced histograms and summary statistics. Here we introduced the Gaussian distribution to motivate the average and standard deviation as useful summary statistics.

***Are health outcomes and income inequality increasing or decreasing in the world?***

This case study is inspired by Hans Rosling's Ted Talk[1] titled 'New insights into poverty' and Jenny Bryan's STAT 545 course[2] on data wrangling. We asked the students if they thought we could divide countries into 'developing' and 'developed' and if income inequality across countries is getting better or worse. To answer these questions we used a subset of the data[3] from Gapminder[4], which provides annual, population-level summaries per country starting in 1810. Examples of population-level summaries in Gapminder for countries include total population, life expectancy at birth and per-capita gross domestic product (GDP). Since this data is provided as a series of comma separated values (CSV) files, we used it to demonstrate the common task of joining data sets together as a form of data wrangling. Here, we introduce several powerful functions in *dplyr* for exactly this purpose. With the data organized, we were ready to answer the original questions. As part of **connecting** the question with the data, students face the challenge of quantifying the term "developed" given the data they have access to. Examining the available summary statistics students realize that fertility rate and life expectancy most relate to the conventional definition that splits the world into small family/long life versus large family/short life. Through the data exploration, such as scatterplots of these summaries by year, they realize that before 1960, there were two very obvious clusters. At this point we introduce several *ggplot2* utilities to allow students to split (or facet) plots by year and label countries by, for example, continent or Organisation for Economic Co-operation and Development (OECD) membership. At this point, students quickly realize that the definition from the 1960s no longer applies since the clusters start coming together with time and are almost completely merged by the present day. The question of income inequality pushes students to connect the shape of a distribution to interpretable properties of the underlying data. For example, in the 1960s income per person revealed two distinct modes. To understand what explains this bimodality, the use of stratification and boxplots quickly demonstrates that OECD countries were much richer than the rest of the world. Finally, by creating histograms of income per person across time they observe that the modes move closer together, which can be interpreted as a reduction of income inequality across countries.

---

[1] http://www.ted.com/talks/hans_rosling_reveals_new_insights_on_poverty
[2] http://stat545.com/

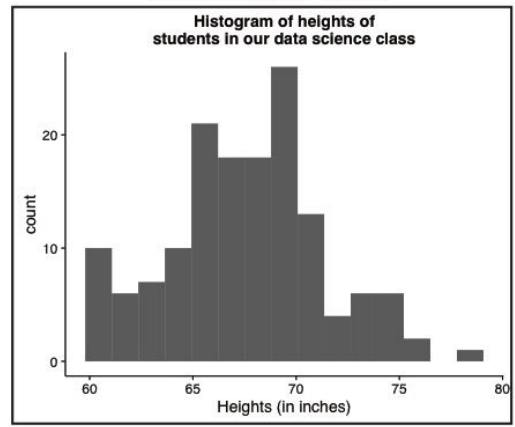

**Supplemental Figure 1.** An example of analyzing one data set as an in-depth case study for several lectures. (A) Students self-reported their heights in inches using a Google Form. (B) The self-reported heights motivated the need and frustration often found in data wrangling. (C) Two lectures were created analyzing the self-reported heights and written in R Markdown combining narrative text and code, making it a reproducible document that students could follow along with using their laptops in class. (D) Exploratory data analysis was introduced as a tool to summarize and identify problems with data.

---

[3] https://github.com/jennybc/gapminder
[4] https://www.gapminder.org/data/

Throughout this case study, we facilitated data manipulation and data visualization using R packages such as *dplyr* and *ggplot2*. We dedicated individual lectures at the beginning of the course to both these packages as they were used throughout the course.

***Who is going to win the 2016 republican primary?***

To answer this question, we start by introducing students to poll aggregators and suggest they imitate their approach. However, to do this, quite a bit of statistical knowledge is necessary. To motivate the students, we ask the questions 'What do the polls report?' and 'What does the margin of error mean?'. As a simplification, we start out by framing the problem as drawing red and blue beads from an urn. Specifically, we ran a competition related to guessing the proportion of red beads in a jar with blue and red beads (Supplemental Figure 2). We built a Shiny application[5] to simulate the experience for a student to take a random sample of at most *N* blue and red beads from the jar with hidden percentage of 52.9% red beads. The goal for the student is to provide an interval around their guess of the proportion of red beads in the jar and the prize goes to the student with the smallest interval containing the true proportion, with ties decided by the smallest sample size *N*. This motivates students to want to learn the concepts of probability, population parameters, estimates, standard errors, and confidence intervals. Students are then able to **connect** statistical inference to polls. Then, we realize that our confidence intervals were not useful in practice as most include 50% and would not permit us to predict a winner. Finally, we demonstrate the power of aggregation as we are able to form a very small confidence interval by combining the submissions from all students. Throughout this part of the course we used Monte Carlo simulations to illustrate probability and inference concepts as proposed by Cobb (2007), Hesterberg (2015) and Horton (2015).

---

[5] https://dgrtwo.shinyapps.io/urn_drawing/

Having introduced the theory, we are ready to **connect** the simplified problem of drawing red beads from an urn to the real-world problem of the 2016 Republican Primary. We use U.S. opinion polls from previous election cycles to empirically determine what statistical theory predicts correctly and what it does not. We use data from the pollstR R package (CRAN pollstR 2016), which provides U.S. opinion polls since 2004 from the Huffington Post Pollster's website. With the use of historical data, we discover pollster variability, temporal changes, and election cycle specific bias. This starts a conversation about how to best aggregate the data and form confidence intervals since sampling variability is not able to fully describe the observed variability. We use this to motivate the concept of statistical models. We ask students to propose models and to describe how they would estimate parameters using historical data. At this point, students are introduced to the possibility of scraping poll data from the web, and we introduce several computing tools and strategies to do this effectively. Finally, we briefly introduce the concept Bayesian inference by asking students what the pollsters mean with their statements about the 'probability of a candidate winning?'.

***How to pick players for a baseball team with a limited budget?***

In this case study, we challenge the students to build a hypothetical baseball team for the 2002 season with a budget of 50 million US$. We motivate the challenge by describing the 2011 movie *Money Ball*, which tells the story of the 2002 Oakland A's, a cash strapped team

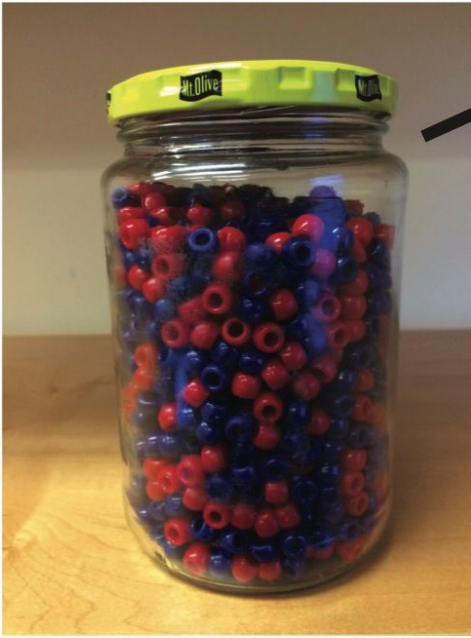
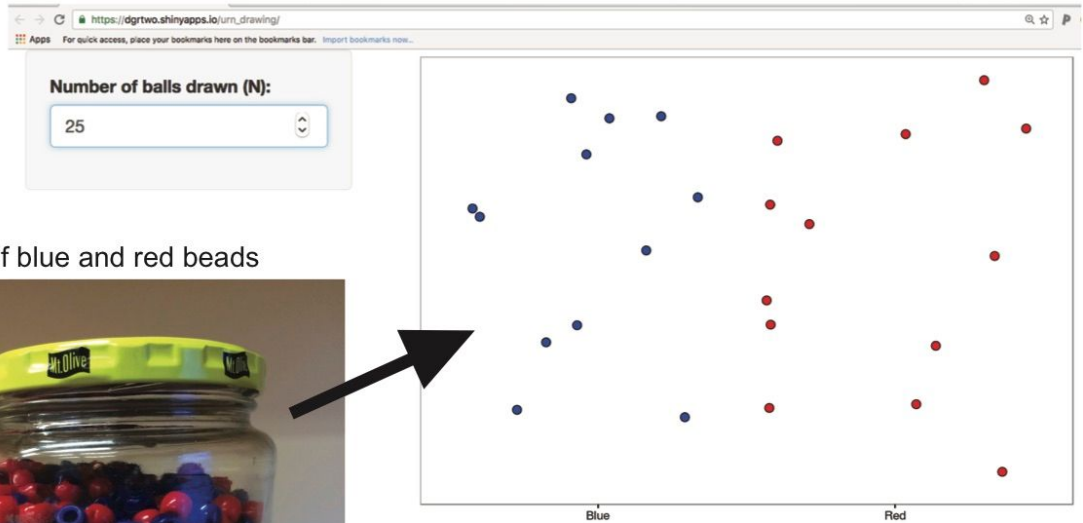

**Supplemental Figure 2.** Using an urn to understand polls. (A) Jar of blue and red beads presented to the class. (B) Shiny application that permitted students to take a random sample from the jar.

that used data analysis to find inefficiencies in the market to build a competitive team. We begin by describing some of the rules of baseball and how they relate to reported statistics, such as 'singles', 'doubles', 'triples', 'home runs', 'stolen bases' and 'bases on balls'. We use data from the Lahman R package (CRAN Lahman 2016), which provides baseball data, including statistics on pitching, hitting and salaries, from the 'Sean Lahman Baseball Database' from 1871 to 2015

(Lahman 2016). Next, we ask students to determine which statistics are most informative to predict runs (as a surrogate for wins). By asking 'How many more runs a team will score if they hit 10 more 'home runs' in a season?', students are lead to **connecting** this real-world problem with linear regression. By asking 'How many more runs a team will score if they draw 10 more 'bases on balls' in a season, students discover that confounding (in this case, players that hit 'home runs' receive more 'bases on balls') leads to an overestimate of runs. By stratifying players by their 'home runs' and re-estimating the effect of 'bases on balls' (conditioned on 'home runs'), we arrive at multiple linear regression. Finally, by examining salary information, we then use similar techniques to determine how much it costs to increase each statistic and which one is more cost-effective. Eventually students arrive at the same conclusion as the Oakland A's did: "buying" 'base on balls' are the best deal.

*How to predict handwritten digits?*

This case study is based on the classic machine learning example: predicting handwritten digits (0, 1, 2, ..., 9) from pixel intensities obtained from a digitized image. We use an adapted dataset from the MNIST Database of Handwritten Digits (LeCun et al. 2016). We begin by simplifying the problem by only considering the '1', '2' and '7' handwritten digits and reducing the feature space from 784 (28 x 28) dimensions to a two-dimensional feature space, where the two predictors represent two interpretable summaries: the proportion of black pixels in the upper left quadrant and lower right quadrant. This enables us to plot the predictor space and motivate the ideas behind different machine learning algorithms. Next, we introduce the concept of conditional probabilities, show the limitation of simple approaches, such as regression and logistic regression (the conditional probabilities are not linear), and motivate the use of techniques based on smoothing, such as k-Nearest Neighbors (kNN) as well as tree-based methods. Then, we return the original feature space (using all 784 features) and introduce machine learning more generally including loss functions, test, training and validation data sets, tuning parameters and cross-validation. We introduce and evaluate the performance of methods to be used for classification, such as naive Bayes, linear discriminant analysis (LDA), quadratic discriminant analysis (QDA), kNN , classification and regression trees (CART), and random forests. Throughout this case study we constantly go back to the two feature example to provide a visual way of understanding how the techniques differ.

***Will I like this movie?***

As an introduction to the challenge of recommender systems, we motivate the students with the problem movie recommendations and describe the Netflix challenge (Bell, Koren and Volinsky 2007). Because the Netflix data is not publicly available, we used movie rating data from MovieLens (Harper and Konstan 2015) after obtaining permission from the authors to use it in our course. We presented the same loss function used in the Netflix challenge and asked the students to consider ways to predict movie ratings. We start with simple models that consider only user and movie effects. For example, using a naive model with only movie effects is intuitive for the students to grasp, because they know that some movies are better than others. We estimate the movie effects by simply taking the average rating of each movie after centering the ratings by removing the grand mean. Next, we motivate the use of regularization by showing that the best movies, according to our naive estimate, are movies with very few ratings. We propose a simple penalized approach to ameliorate this effect. The students discover that if we consider just one movie, the regularized movie effect, $b$, with penalty parameter $\lambda$, can be easily calculated and is given by:

```
b_hat = sum(Y - mean(Y) ) / (lambda + length(Y))
```

or

$$\hat{b}(\lambda) = \frac{1}{\lambda + n} \sum_{u=1}^{n} (Y_u - \hat{\mu})$$

Where the vector Y contains the $Y_u$ which are the movie ratings for the $u^{th}$ user ($u = 1, ..., n$). The students quickly see how this improves our estimates as the students see how simply using $\lambda = 3$, brings some very well known blockbusters to the top of the best movies lists. We follow this up by showing how regularization improves the loss function and gets us closer to winning the improvement needed to win the Netflix prize. Then, we consider adding user effects. Here, the students are a bit more surprised to see that there is large variation across users as some are grumpier and others happier raters. We discover this by data exploration. Once again the students see a way to improve our recommendation system's performance.

Finally, we motivate the need for interaction effects by showing, for example, that users with a preference for *The Godfather* correlate strongly with users with a preference for *Good Fellas,* but correlate negatively with, for example, *Ghost* and *Pretty Woman*. Students quickly realize that there are different clusters of movies and users. This serves as motivation to introduce principal component analysis and matrix factorization. Once we estimate interaction effects with matrix factorization we get an improvement that gets relatively close to the winning entry in the Netflix challenge.

*Is there gender bias in grant funding?*

Here, we begin by sharing the results of a recent PNAS paper (Lee and Ellemers 2015) that analyzed success rates from funding agencies in The Netherlands and concluded the data 'reveal gender bias favoring male applicants over female applicants in the prioritization of their

"quality of researcher" (but not "quality of proposal") evaluations and success rates'. We present the data used in the paper with the students and challenge them to discuss if they find the results convincing or if they are skeptical of the results Next, we describe the famous 1973 admission data from U.C. Berkeley, which is a clear example of how Simpson's Paradox resulted in incorrect conclusions. We encourage the students to explore the admissions data and they quickly notice that once you stratify the data by subject the apparent gender bias disappears. Finally, we return to the example from The Netherlands in which the students similarly stratify the data, but this time they stratify by research area and they realize that this too is a case of Simpson's paradox (Volker and Steenbeek 2015).

# Supplemental References

American Statistical Association Undergraduate Guidelines Workgroup. (2014), "Curriculum Guidelines for Undergraduate Programs in Statistical Science," *Nov 2016*, Available at *https://www.amstat.org/asa/files/pdfs/EDU-guidelines2014-11-15.pdf*.

CRAN Lahman Package. (2016), Available at *https://cran.r-project.org/web/packages/Lahman/*.

CRAN PollstR Package. (2016), Available at *https://cran.r-project.org/web/packages/pollstR/*.

Donoho, D. (2015), "50 years of Data Science", Paper presented at the *Tukey Centennial Workshop* on Sept 18, 2015. Princeton, NJ.

Harper, F. M., Konstan, J. A. (2015), "The MovieLens Datasets: History and Context," *ACM Transactions on Interactive Intelligent Systems* 5, Article 19.

Horton, N. J., Hardin, J.S. (2015), "Teaching the Next Generation of Statistics Students to 'Think With Data': Special Issue on Statistics and the Undergraduate Curriculum", *The American Statistician*, 69, 259–65.

Khachatryan, D. (2015), "Incorporating Statistical Consulting Case Studies in Introductory Time Series Courses," *The American Statistician*, 69, 387–96.

Lahman, S. 2016. Sean Lahman's Baseball Database. Available at *http://www.seanlahman.com/baseball-archive/statistics/*

LeCun, Y., Cortes, C., Burges, C. J. C. (2016), The MNIST Database of Handwritten Digits, Available at *http://yann.lecun.com/exdb/mnist*

Lee, R. V. D., Ellemers, N. (2015), "Gender Contributes to Personal Research Funding Success in The Netherlands," *Proceedings of the National Academy of Sciences*, 112, 12349-2353.

Nolan, D., Temple Lang, D. (2015) "Explorations in Statistics Research: An Approach to Expose Undergraduates to Authentic Data Analysis," *The American Statistician*, 69, 292–99.

Smucker, B. J., Bailer, A. J. (2015), "Beyond Normal: Preparing Undergraduates for the Work Force in a Statistical Consulting Capstone," *The American Statistician*, 69, 300–306.

Tukey, J.W. (1962) "The future of data analysis," *The Annals of Mathematical Statistics* 33, 1-67.

Volker, B., Steenbeek, W. (2015), "No Evidence That Gender Contributes to Personal Research Funding Success in The Netherlands: A Reaction to Van Der Lee and Ellemers," *Proceedings of the National Academy of Sciences*, 112, E7036–E7037.

Wu, J. (1998), "Identity of statistics in science examined". *The University Record*. Talk presented '*Statistics = Data Science?*' at University of Michigan on November 10, 1998. Available at http://ur.umich.edu/9899/Nov09_98/4.htm.